%% file: main.tex
\documentclass[sigconf]{acmart}
\AtBeginDocument{%
  }

\usepackage{algorithm}
\usepackage{algorithmic}
\usepackage{multirow}
\usepackage[absolute,overlay]{textpos}
\setcopyright{acmlicensed}
\copyrightyear{2018}
\acmYear{2018}
\acmDOI{XXXXXXX.XXXXXXX}
\acmConference[Conference acronym 'XX]{Make sure to enter the correct
  conference title from your rights confirmation email}{June 03--05,
  2018}{Woodstock, NY}
\acmISBN{978-1-4503-XXXX-X/2018/06}

\begin{document}

\begin{textblock*}{\paperwidth}(2.0cm, 1.5cm)\includegraphics[height=1.3cm]{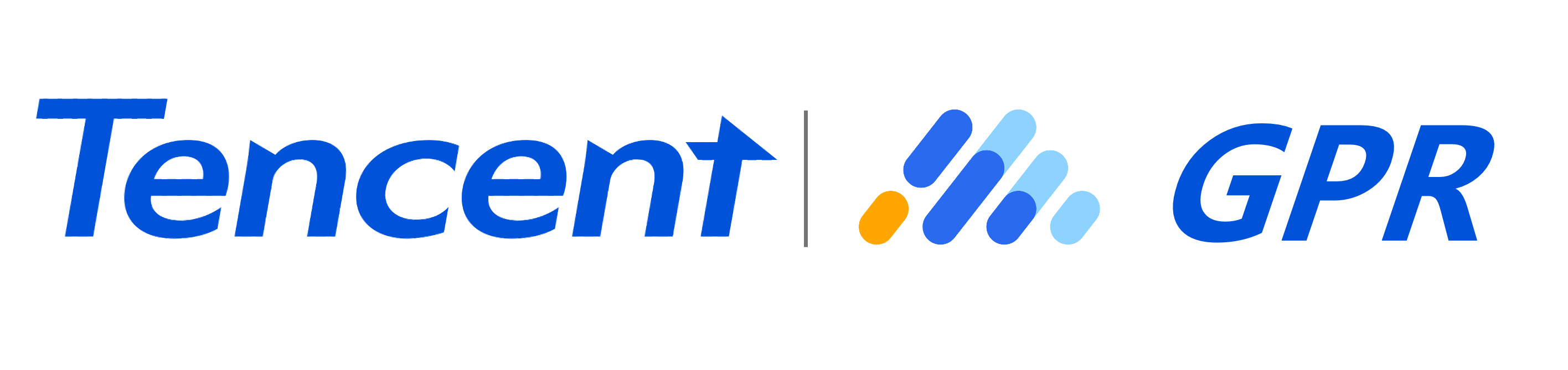} \end{textblock*}

\title{DiffuReason: Bridging Latent Reasoning and Generative Refinement for Sequential Recommendation}


\author{Jie Jiang}
\authornote{Both authors contributed equally to this research.} 
\affiliation{%
  \institution{Tencent}
    \state{Beijing}
  \country{China}
}
\email{zeus@tencent.com}

\author{Yang Wu}
\authornotemark[1]
\affiliation{%
  \institution{Tencent}
  \state{Beijing}
  \country{China}
}
\email{samuelywu@tencent.com}

\author{Qian Li}
\authornotemark[1]
\affiliation{%
  \institution{Tencent}
  \state{Beijing}
  \country{China}
}
\email{kathieqli@tencent.com}

\author{Yuling Xiong}
\authornotemark[1]
\affiliation{%
  \institution{Tencent}
  \state{Beijing}
  \country{China}
}
\email{whitnyxiong@tencent.com}

\author{Yihang Su}
\affiliation{%
  \institution{Tencent}
    \state{Beijing}
  \country{China}
}
\email{charliesu@tencent.com}

\author{Junbang Huo}
\affiliation{%
  \institution{Tencent}
    \state{Beijing}
  \country{China}
}
\email{remoohuo@tencent.com}

\author{Longfei Lu}
\affiliation{%
  \institution{Tencent}
    \state{Beijing}
  \country{China}
}
\email{loneffylu@tencent.com}

\author{Jun Zhang}
\authornote{Corresponding author.}
\affiliation{%
  \institution{Tencent}
    \state{Beijing}
  \country{China}
}
\email{neoxzhang@tencent.com}

\author{Huan Yu}
\affiliation{%
  \institution{Tencent}
    \state{Beijing}
  \country{China}
}
\email{	huanyu@tencent.com}

\renewcommand{\shortauthors}{Qian Li}

\begin{abstract}

Latent reasoning has emerged as a promising paradigm for sequential recommendation, enabling models to capture complex user intent through multi-step deliberation. Yet existing approaches often rely on deterministic latent chains that accumulate noise and overlook the uncertainty inherent in user intent, and they are typically trained in staged pipelines that hinder joint optimization and exploration. To address these challenges, we propose DiffuReason, a unified “Think-then-Diffuse” framework for sequential recommendation. It integrates multi-step Thinking Tokens for latent reasoning, diffusion-based refinement for denoising intermediate representations, and end-to-end Group Relative Policy Optimization (GRPO) alignment to optimize for ranking performance. In the Think stage, the model generates Thinking Tokens that reason over user history to form an initial intent hypothesis. In the Diffuse stage, rather than treating this hypothesis as the final output, we refine it through a diffusion process that models user intent as a probabilistic distribution, providing iterative denoising against reasoning noise. Finally, GRPO-based reinforcement learning enables the reasoning and refinement modules to co-evolve throughout training, without the constraints of staged optimization. Extensive experiments on four benchmarks demonstrate that DiffuReason consistently improves diverse backbone architectures. Online A/B tests on a large-scale industrial platform further validate its practical effectiveness.

\end{abstract}


\ccsdesc[500]{Information systems ~ Recommender systems}

\keywords{Recommender Systems, Large Language Models, Chain-of-Thought, Latent Reasoning}

\received{20 February 2007}
\received[revised]{12 March 2009}
\received[accepted]{5 June 2009}

\maketitle

\input{paragraphs/1.Introduction}
\input{paragraphs/2.related}

\input{paragraphs/3.Methods}
\input{paragraphs/4.Experiments}

\bibliographystyle{ACM-Reference-Format}
\bibliography{sample-base}

\appendix
\input{paragraphs/5.Appendix}

\end{document}

%% file: paragraphs/1.Introduction.tex
\section{Introduction}~\label{sec:intro}

Sequential recommendation systems~\cite{wang2019sequential, fang2020deep} serve as pivotal infrastructure in modern digital platforms, facilitating personalized discovery across diverse domains such as e-commerce~\cite{linden2003amazon, chen2019behavior}, short-video streaming~\cite{liu2019user, gong2022real}, and social media~\cite{ konstas2009social, yang2012circle}. Recently, the field has witnessed a growing interest from intuitive "Fast Thinking" (System 1) to deliberative "Slow Thinking" (System 2)~\cite{kahneman2011thinking}. Specifically, System 2 refers to the mechanism where the model performs multi-round reasoning to extend its computational depth~\cite{wei2022chain, yang2025llm2, zhang2025system}, thereby empowering it to effectively capture complex user intents~\cite{tsai2024leveraging, zhang2025slow, fang2025reason4rec, zhao2025reason}. Consequently, investigating how to harness this slow thinking capability to achieve more precise reasoning has become a focal point of current research.

In pursuit of this goal, prior studies~\cite{tang2025think,zhang2025reinforced, zhang2026cross, tang2026parallel} have made significant strides, validating the effectiveness of latent reasoning in enhancing recommendation performance. However, these methods face a critical challenge due to the lack of high-quality supervision signals~\cite{guo2026s}. Since ground-truth reasoning paths are unavailable, existing approaches predominantly rely on proxy losses (e.g., reconstruction objectives~\cite{xing2025reg4rec}), which fail to effectively filter out reasoning noise. Furthermore, these methods are typically constrained by a deterministic generation paradigm. This limitation makes them ill-equipped to model the inherent uncertainty of user intents, as they are restricted to a static reasoning trajectory.

Recent efforts~\cite{liu2025lares,zhang2025reinforced, kong2025minionerec, zhang2026reasoning} have sought to bridge this supervision gap by integrating Reinforcement Learning (RL)~\cite{sutton1998reinforcement} to align reasoning directly with recommendation objectives . While this represents a promising direction, these approaches predominantly operate under a disjointed "Two-Stage" training paradigm, typically consisting of a warm-up pre-training phase followed by RL fine-tuning. This sequential separation poses a fundamental optimization barrier. Specifically, the initial stage inevitably biases the policy towards specific trajectories, thereby constraining the exploration potential of the subsequent RL optimization~\cite{chaudhari2025rlhf, su2025trust}. Consequently, this decoupling restricts the synergistic evolution of the reasoning module and the recommendation policy, hindering end-to-end alignment.

\begin{figure*}[t]
    \centering
    \includegraphics[width=1\textwidth]{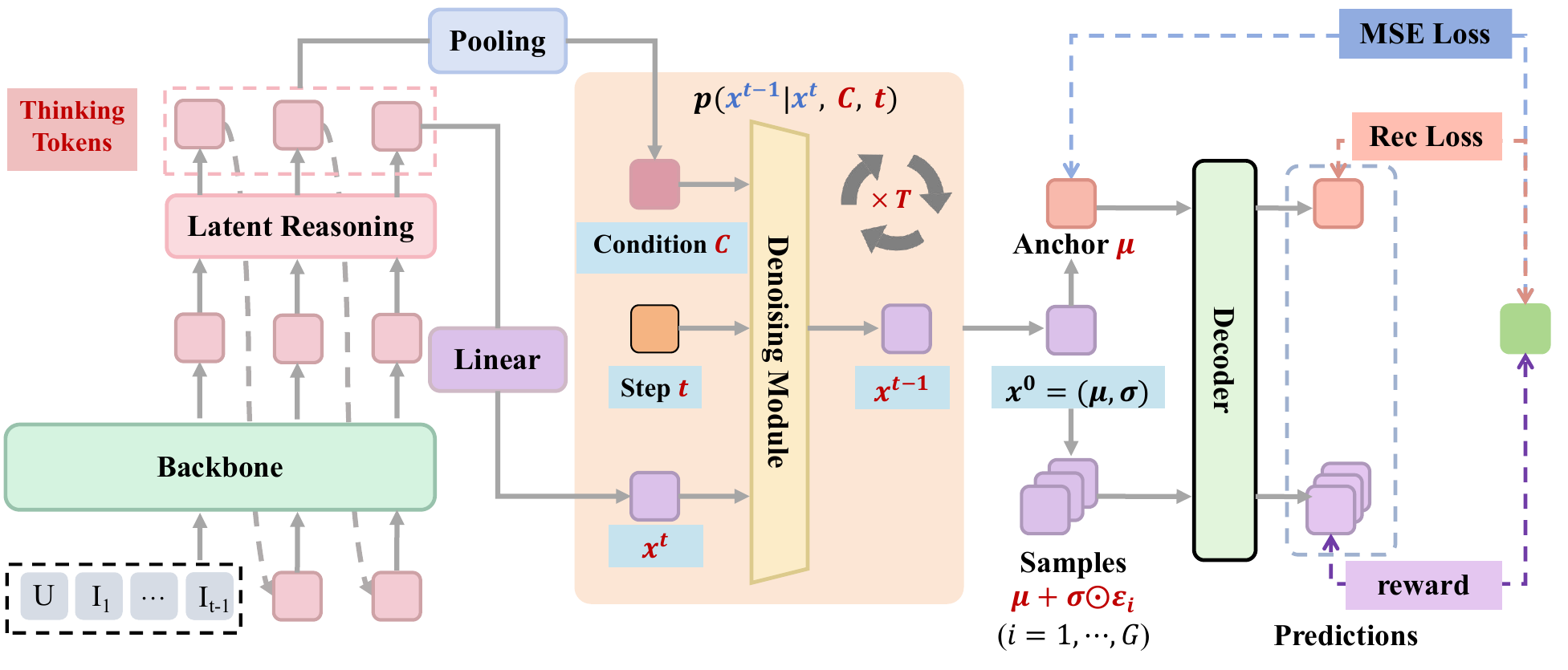} 
\caption{The overall architecture of the proposed method.  The Latent Reasoning module generates explicit Thinking Tokens, which are aggregated via Pooling to form the Condition $c$. (Middle) The Refining Module, driven by a Denoising Module, iteratively recovers the latent state $x^0$ to produce a deterministic Anchor $\mu$ and stochastic samples. The Decoder maps these Refining Tokens to predictions, which are optimized jointly by the recommend loss (Rec Loss) and the alignment reward.}
    \label{fig:framework}
\end{figure*}

In this work, we propose DiffuReason, a unified framework designed to mitigate reasoning noise and bridge the optimization gap. Moving beyond rigid "one-pass" generation, DiffuReason follows a simple “Think-then-Diffuse” paradigm. Specifically, it leverages Diffusion Models to treat latent reasoning as a probabilistic condition rather than a fixed result. This enables an iterative denoising process that refines coarse thoughts into more precise user representations. Crucially, to ensure this refinement is driven by the ultimate goal, we replace the disjointed two-stage paradigm with an end-to-end training strategy. By integrating recommendation rewards directly into the optimization, our framework allows the diffusion process to be explicitly guided by ranking metrics. This fosters the unified optimization of the backbone, reasoning, and refinement modules, aligning the entire system with the recommendation objective. We further validate DiffuReason through online A/B testing on a large-scale industrial platform, demonstrating its practical effectiveness in real-world production.

In summary, our contributions are as follows:

\begin{itemize}
    \item We propose the ``Think-then-Diffuse" approach to enhance the robustness of latent reasoning. Instead of relying on deterministic reasoning paths, we treat latent reasoning as a generative condition and leverage Diffusion models to iteratively refine coarse reasoning steps, effectively mitigating the noise and uncertainty inherent in latent reasoning.
    
    \item We establish an end-to-end training framework that bridges the gap between generative reasoning and ranking objectives. This method enables the diffusion refiner to be optimized directly against final recommendation metrics, ensuring explicit alignment between the generation process and recommendation accuracy.
    
    \item We conduct extensive experiments on several benchmark datasets, consistently outperforming all baselines. Crucially, our method is model-agnostic, yielding significant improvements  across various backbone architectures. We further validate its practical effectiveness via online A/B testing on a large-scale industrial platform.
\end{itemize}

%% file: paragraphs/2.related.tex
\section{Related Work}

\textbf{Reasoning-Enhanced Recommendation.}
Recent studies explore integrating reasoning capabilities into recommender systems to capture complex user interests. Early approaches~\cite{long2024got4rec, fang2025reason4rec, wu2025scoter, liu2025improving, yi2025recgpt, yu2025thinkrec, tang2025coder} utilize Large Language Models to generate explicit textual rationales for embedding augmentation, but they suffer from high latency and resource consumption. Consequently, the focus has shifted to implicit reasoning~\cite{
tang2025think,zhang2025reinforced, zhang2025slow, zhang2026cross, tang2026parallel}, where models like ReaRec~\cite{tang2025think} and LARES~\cite{liu2025lares} perform inference-time deduction directly in the latent space. While efficient, supervising these implicit thoughts remains a critical challenge~\cite{guo2026s}. Existing methods typically rely on heuristic proxies (e.g., step-level alignment) or disjointed two-stage training, which fails to align the reasoning process with the final ranking objective. DiffuReason addresses this by introducing an end-to-end reinforcement learning framework that directly optimizes the reasoning-refinement chain for recommendation performance.

\textbf{Diffusion Models for Recommendation.}
Diffusion Models~\cite{ho2020denoising} have shown great promise in recommendation for modeling uncertainty and denoising interaction sequences~\cite{li2023diffurec, yang2023generate, liu2023diffusion, zhao2024denoising, di2025federated, li2025dimerec, wei2025diffusion}. By iteratively refining noise into structured representations, they capture multi-modal user interests more effectively than deterministic models. However, a fundamental limitation of current diffusion-based recommenders is their reliance on Mean Squared Error loss for reconstruction. This point-wise objective is often misaligned with the list-wise ranking goal required for top-k recommendation~\cite{liu2024preference}. We bridge this gap by treating diffusion as a refinement module rather than a standalone generator and employing a GRPO-based~\cite{shao2024deepseekmath} alignment strategy to ensure the generative process directly translates to superior ranking metrics.

%% file: paragraphs/3.Methods.tex
\section{Method}
\label{sec:method}

We present DiffuReason, a unified framework that synergizes latent reasoning with generative refinement. As illustrated in Figure \ref{fig:framework}, our method addresses the sparsity and noise in user behaviors through a ``Think-then-Diffuse'' paradigm, progressing in three stages: (1) Reasoning, which generates Thinking Tokens to capture multi-step intents; (2) Refining, which employs a diffusion process to model the inherent uncertainty in user interests; and (3) Alignment, which aligns the generation process with recommendation objectives.

\subsection{Preliminaries and Background}
\label{sec:preliminaries}

\paragraph{Task Definition.}
Let $\mathcal{V}$ denote the set of all items. For a user $u$, the historical interaction sequence is represented as $S_u = [v_1, v_2, \dots, v_L]$, where $v_i \in \mathcal{V}$ is the item interacted with at step $i$ and $L$ is the sequence length. The objective of sequential recommendation is to predict the probability of the next item $v_{L+1}$ given the history $S_u$. For notation simplicity, we denote the ground-truth target item (and its corresponding embedding vector) as $v_{target}$.

\paragraph{Item Representation.}
To ensure generality, we consider two distinct paradigms for representing items. The first is ID-based Representation (e.g., SASRec~\cite{kang2018self}), which assigns a unique, independent embedding vector to each item. This approach treats items as discrete nodes, often ignoring intrinsic semantic similarities. The second is SID-based Representation (e.g., TIGER~\cite{rajput2023recommender}), which represents items as sequences of Semantic IDs. These discrete codes are generated by quantizing item features (e.g., via RQ-VAE~\cite{zeghidour2021soundstream, lee2022autoregressive}), allowing the model to capture content structures through shared code prefixes.

\paragraph{Backbone Encoder.}
Regardless of the representation, we utilize a generic backbone $\mathcal{F}_\theta$ to project the interaction history into a sequence of hidden states $H = [h_1, \dots, h_{L}]$. This output serves as the universal context for our method.

\subsection{Latent Reasoning with Thinking Tokens}
\label{sec:step1}

Standard backbones typically rely on the final state for immediate prediction, akin to fast, intuitive thinking (System 1). This often fails to capture complex, multi-hop user intents. To overcome this, we introduce a latent reasoning mechanism that simulates slow, deliberative thinking (System 2). We generate a sequence of Thinking Tokens $\mathcal{T} = \{\tau_1, \dots, \tau_R\}$ to model the deduction process. Initialized with the current state $\tau_1 = h_{L}$, the tokens evolve auto-regressively:
\begin{equation}
    \tau_r = \text{MLP}(\text{Attention}(\tau_{r-1}, H)), \quad r=2,\dots,R
    \label{eq:reasoning}
\end{equation}
This recursive formulation allows the model to iteratively attend to historical contexts, uncovering deep dependencies that are unobservable in a single forward pass.

To translate this discrete reasoning chain into a semantic guide for the generative phase, we aggregate the tokens using a generic function $\psi(\cdot)$ (e.g., Attention Pooling) to derive the unified Condition Vector $c$:
\begin{equation}
\label{eq:condition}
    c = \psi(\mathcal{T})
\end{equation}
While this vector $c$ encapsulates the deliberate reasoning, it remains deterministic and may still harbor noise derived from sparse inputs. This limitation necessitates the subsequent refinement stage to model the inherent uncertainty.

\subsection{Generative Refinement via Diffusion}
\label{sec:diffusion}

The reasoning context derived in the previous stage acts as a semantic guide. However, relying solely on deterministic reasoning captures only a static snapshot of user intent. To bridge this gap, we introduce a generative refinement module that explicitly models the uncertainty in user interests.  We provide the full background on
diffusion models in Appendix~\ref{app:diffusion}.

Instead of generating representations from random noise, we leverage the reasoning output as a warm start. Specifically, we initialize the diffusion trajectory by injecting noise into the last thinking token $\tau_R$:
\begin{equation}
    x^T = \tau_R + \epsilon_{init}, \quad \epsilon_{init} \sim \mathcal{N}(0, I)
    \label{eq:init_noise}
\end{equation}

Simultaneously, to preserve global context, we utilize the vector $c$ (Eq. \ref{eq:condition}) as the condition.

To implement the reverse transition, we employ a Multi-Layer Perceptron (MLP) as the denoising network. This network takes the current noisy state $x^t$, the timestep embedding $t$, and the condition $c$ as inputs to predict the distribution parameters of the previous latent state. Mathematically, the reverse process is modeled as a Gaussian transition parameterized by these MLP outputs:
\begin{equation}
    p_\phi(x^{t-1} | x^t, c) = \mathcal{N}(x^{t-1}; \boldsymbol{\mu}_\phi(x^t, t, c), \boldsymbol{\sigma}_\phi(x^t, t, c))
    \label{eq:denoise}
\end{equation}
By iteratively applying this denoising step, the model progressively corrects the coarse estimation from $\tau_R$ into a refined representation.

After $T$ denoising steps, the module produces the final latent state $x^0$, which parameterizes a Gaussian distribution with mean $\boldsymbol{\mu}$ and diagonal standard deviation $\boldsymbol{\sigma}$:
\begin{equation}
    x^0 = (\boldsymbol{\mu}, \boldsymbol{\sigma})
    \label{eq:final_state}
\end{equation}
Here, $\boldsymbol{\mu}$ serves as the stable center of the refined intent, while $\boldsymbol{\sigma}$ captures the learned uncertainty.

To fully exploit this structure, we generate a set of candidate representations to be sent to the decoder. We define the primary prediction as the stable anchor $z_{anchor} = \boldsymbol{\mu}$. Additionally, we sample $G$ stochastic variations from the distribution to capture diverse intents:

\begin{equation}
\label{eq:sampling}
    z_i = \boldsymbol{\mu} + \boldsymbol{\sigma} \odot \epsilon_i, \quad z_i \sim \mathcal{N}(\boldsymbol{\mu},\, \mathrm{diag}(\boldsymbol{\sigma}^2)), \quad i=1, \dots, G
\end{equation}
where $\epsilon_i \sim \mathcal{N}(0, I)$. Finally, these vectors are fed into the decoder for prediction. This design allows the model to output both a high-confidence main prediction via $z_{anchor}$ and diverse exploratory options via the samples.

During the joint training, we apply a standard reconstruction loss to the diffusion module. Specifically, we enforce the predicted Anchor State $\boldsymbol{\mu}$ to align with the ground-truth target item embedding $v_{target}$ via Mean Squared Error (MSE):

\begin{equation}
    \mathcal{L}_{diff} = \| \boldsymbol{\mu} - v_{target} \|^2
    \label{eq:loss_diff}
\end{equation}

While this continuous loss ensures the model recovers the target semantics, it acts only as a proxy. Minimizing Euclidean distance does not guarantee the correct ordering of candidates. This discrepancy motivates the end-to-end alignment strategy introduced in the next section.

\subsection{End-to-End Alignment via GRPO}
\label{sec:alignment}

To effectively bridge the gap between the generative reconstruction objective and the discrete ranking metric, we employ the Group Relative Policy Optimization (GRPO) algorithm. GRPO optimizes the policy by leveraging group-wise comparisons, eliminating the need for a separate value network. Specifically, for each input sequence $S_u$, we sample a group of $G$ independent noise trajectories $\{\epsilon_i\}_{i=1}^G$ from the current policy. By applying the sampling mechanism defined in Eq. \ref{eq:sampling}, we obtain the corresponding set of candidate representations $\{z_1, \dots, z_G\}$. These sampled variations explore the neighborhood around the Anchor State, ensuring refinements stay close to the reasoned intent.

Subsequently, to strictly enforce the recommendation goal, we map each latent candidate back to the discrete item space using a generalized decoding function $Decoder(\cdot)$. The top-1 predicted item is obtained via:
\begin{equation}
    \hat{v}_{1,i} = \text{Decoder}(z_i)
    \label{eq:decoder}
\end{equation}
It is worth noting that the specific architecture of the Decoder adapts to the backbone model. For instance, in classic ID-based models, this typically involves a projection layer followed by Softmax; whereas for semantic-ID or generative backbones, it entails a hierarchical or autoregressive decoding process.

Based on these predictions, a discrete binary reward $r_i$ is assigned using the Hit@1 metric:
\begin{equation}
    r_i = \mathbb{I}(\hat{v}_{1,i} = v_{target})
    \label{eq:reward}
\end{equation}
where $\mathbb{I}(\cdot)$ is the indicator function. This binary signal directly penalizes or encourages the model based on whether the generated refinement leads to the correct item.

With these rewards, the advantage $A_i$ is derived by normalizing scores within the sampled group:
\begin{equation}
    A_i = \frac{r_i - \text{Mean}(\{r_j\}_{j=1}^G)}{\text{Std}(\{r_j\}_{j=1}^G) + \epsilon_{adv}}
    \label{eq:advantage}
\end{equation}
This formulation stabilizes the training process by comparing each trajectory against its peers. Finally, the alignment parameters are updated by maximizing the objective involving the clipped importance sampling ratio $\rho_i$ 
(derived in Appendix~\ref{app:grpo}):
\begin{equation}
\resizebox{0.91\hsize}{!}{$
    \mathcal{L}_{align} = - \frac{1}{G} \sum_{i=1}^G \min \left( \rho_i A_i, \text{clip}(\rho_i, 1-\epsilon_{clip}, 1+\epsilon_{clip}) A_i \right)
$}
\label{eq:loss_align}
\end{equation}
Notably, we exclude the explicit KL divergence term typically found in standard GRPO. Since our refinement module operates as a residual generator centered around the deterministic Anchor State, the exploration is intrinsically constrained within the local neighborhood of the reasoned intent. This structural regularization renders the additional KL penalty redundant.

Finally, our framework integrates these objectives through an end-to-end training strategy that simultaneously targets accurate next-item prediction, stable latent reconstruction, and ranking-aware alignment. The total objective function is defined as a weighted combination of these components:
\begin{equation}
    \mathcal{L}_{total} = \mathcal{L}_{rec} + \alpha \mathcal{L}_{diff} + \beta \mathcal{L}_{align}
    \label{eq:loss_total}
\end{equation}
Here, $\mathcal{L}_{rec}$ is the standard cross-entropy loss applied to the anchor state to ensure basic recommendation accuracy:
\begin{equation}
    \mathcal{L}_{rec} = - \log P(v_{target} | z_{anchor})
    \label{eq:loss_rec}
\end{equation}
$\alpha$ and $\beta$ are hyperparameters that balance the generative consistency and the alignment constraints within the unified optimization landscape.

\begin{algorithm}[tb]
   \caption{End-to-End Training of DiffuReason}
   \label{alg:training}
\begin{algorithmic}[1]
   \STATE {\bfseries Input :} Dataset $\mathcal{D}$, Group size $G$, Hyperparameters $\alpha, \beta, \eta$
   \STATE {\bfseries Output :} Optimized Model parameters $\theta$
   \STATE {\bfseries Initialize :} Model parameters $\theta$
   
   \REPEAT
       \STATE Sample batch $(S_u, v_{target}) \sim \mathcal{D}$
       
       \STATE \textcolor{blue}{/* \hspace{0.2em} Stage 1: Latent Reasoning \hfill */}
       \STATE $H \leftarrow \text{Backbone}(S_u); \quad \tau_1 \leftarrow h_L$
       \FOR{$r=2$ {\bfseries to} $R$}
           \STATE $\tau_r \leftarrow \text{MLP}(\text{Attn}(\tau_{r-1}, H))$ \hfill \textcolor{blue}{$\triangleright$ Eq. \ref{eq:reasoning}}
       \ENDFOR
       \STATE $c \leftarrow \text{Pooling}(\mathcal{T}=\{\tau_1, \dots, \tau_R\})$ \hfill \textcolor{blue}{$\triangleright$ Eq. \ref{eq:condition}}
       \STATE $x^T \leftarrow \tau_R + \epsilon_{init}, \quad \text{where } \epsilon_{init} \sim \mathcal{N}(0, I)$ \hfill \textcolor{blue}{$\triangleright$ Eq. \ref{eq:init_noise}}
       
       \STATE \textcolor{blue}{/* \hspace{0.2em} Stage 2: Generative Refinement \hfill */}
       \FOR{$t=T$ {\bfseries down to} $1$}
           \STATE $(\boldsymbol{\mu}_\phi, \boldsymbol{\sigma}_\phi) \leftarrow \text{DenoiseNet}(x^t, t, c)$ \hfill \textcolor{blue}{$\triangleright$ Eq. \ref{eq:denoise}}
           \STATE Sample $x^{t-1} \sim \mathcal{N}(\boldsymbol{\mu}_\phi, \boldsymbol{\sigma}_\phi)$
       \ENDFOR
       \STATE $(\boldsymbol{\mu}, \boldsymbol{\sigma}) \leftarrow (\boldsymbol{\mu}_\phi, \boldsymbol{\sigma}_\phi)$ \hfill \textcolor{blue}{$\triangleright$ Eq. \ref{eq:final_state}}
       \STATE $z_{anchor} \leftarrow \boldsymbol{\mu}$
       \STATE $\mathcal{L}_{diff} = \| z_{anchor} - v_{target} \|^2$ \hfill \textcolor{blue}{$\triangleright$ Eq. \ref{eq:loss_diff}}
       \STATE $\mathcal{L}_{rec} = - \log P(v_{target} | z_{anchor})$  \hfill \textcolor{blue}{$\triangleright$ Eq. \ref{eq:loss_rec}}
       
       \STATE \textcolor{blue}{/* \hspace{0.2em} Stage 3: Alignment via GRPO \hfill */}
       \STATE Sample noise $\{\epsilon_i\}_{i=1}^G \sim \mathcal{N}(0, I)$
       \STATE Obtain candidates $\{z_i\}_{i=1}^G$ via Eq. \ref{eq:sampling} \hfill \textcolor{blue}{$\triangleright$ Eq. \ref{eq:sampling}}
       \FOR{$i=1$ {\bfseries to} $G$}
           \STATE $r_i = \mathbb{I}(\text{Decoder}(z_i) = v_{target})$ \hfill \textcolor{blue}{$\triangleright$ Eq. \ref{eq:reward}}
       \ENDFOR
       \STATE Compute normalized advantages $A_i$ \hfill \textcolor{blue}{$\triangleright$ Eq. \ref{eq:advantage}}
       \STATE Compute ratio $\rho_i$ and $\mathcal{L}_{align}$ using $A_i$ \hfill \textcolor{blue}{$\triangleright$ Eq. \ref{eq:loss_align}}
       
       \STATE \textcolor{blue}{/* \hspace{0.2em} Joint Optimization \hfill */}
       \STATE $\mathcal{L}_{total} = \mathcal{L}_{rec} + \alpha \mathcal{L}_{diff} + \beta \mathcal{L}_{align}$ \hfill \textcolor{blue}{$\triangleright$ Eq. \ref{eq:loss_total}}
       \STATE Update $\theta \leftarrow \theta - \eta \nabla_\theta \mathcal{L}_{total}$
   \UNTIL{convergence}
\end{algorithmic}
\end{algorithm}

%% file: paragraphs/4.Experiments.tex
\begin{table}[h]
    \centering
    \caption{Dataset statistics of the evaluation benchmarks. "AvgLen" represents the average length of item sequences.}
    \label{tab:datasets}
    \begin{tabular}{lcccc}
    \toprule
    Dataset & \#Users & \#Items & \#Interactions & AvgLen \\
    \midrule
    Beauty & 22,363 & 12,101 & 198,502 & 8.88 \\
    Instruments & 24,772 & 9,922 & 206,153 & 8.32 \\
    Sports & 35,598 & 18,357 & 296,337 & 8.32 \\
    Video \&  Games & 67,658 & 25,535 & 654,867 & 9.68 \\
    \bottomrule
    \end{tabular}
\end{table}

\begin{table*}[ht]
    \centering
    \caption{Performance comparison on four datasets. \textbf{DiffuReason-S/T/H} denote the proposed method with SASRec, TIGER, and HSTU backbones. Best results are \textbf{bolded}, second-best are \underline{underlined}. }
    \label{tab:main_results}
    \small
    \setlength{\tabcolsep}{2pt}
    \begin{tabular}{ll|cc|ccc|ccc|ccc}
    \toprule
    \textbf{Dataset} & \textbf{Metric} & \textbf{ReaRec} & \textbf{LARES} & \textbf{SASRec} & \textbf{DiffuReason-S} & \textbf{Improv.} & \textbf{TIGER} & \textbf{DiffuReason-T} & \textbf{Improv.} & \textbf{HSTU} & \textbf{DiffuReason-H} & \textbf{Improv.} \\
    \midrule
    \multirow{4}{*}{\textbf{Beauty}} 
      & R@5  & 0.0439 & 0.0428 & 0.0387 & 0.0463 & 19.64\% & 0.0392 & 0.0412 & 5.10\% & \underline{0.0567} & \textbf{0.0702} & 23.81\% \\
      & R@10 & 0.0674 & 0.0572 & 0.0605 & 0.0711 & 17.52\% & 0.0594 & 0.0645 & 8.59\% & \underline{0.0930} & \textbf{0.1117} & 20.11\% \\
      & N@5  & 0.0278 & 0.0281 & 0.0249 & 0.0273 & 9.64\%  & 0.0257 & 0.0271 & 5.45\% & \underline{0.0376} & \textbf{0.0456} & 21.28\% \\
      & N@10 & 0.0353 & 0.0325 & 0.0318 & 0.0357 & 12.26\% & 0.0321 & 0.0345 & 7.48\% & \underline{0.0502} & \textbf{0.0595} & 18.53\% \\
    \midrule
    \multirow{4}{*}{\textbf{Instruments}} 
      & R@5  & 0.0867 & 0.0638 & 0.0857 & 0.0889 & 3.73\% & 0.0865 & 0.0910 & 5.20\% & \underline{0.1009} & \textbf{0.1096} & 8.62\% \\
      & R@10 & 0.1089 & 0.0791 & 0.1083 & 0.1174 & 8.40\% & 0.1062 & 0.1121 & 5.56\% & \underline{0.1394} & \textbf{0.1521} & 9.11\% \\
      & N@5  & 0.0734 & 0.0456 & 0.0715 & \underline{0.0753} & 5.31\% & 0.0736 & \textbf{0.0770} & 4.62\% & 0.0635 & 0.0725 & 14.17\% \\
      & N@10 & 0.0807 & 0.0505 & 0.0788 & \underline{0.0845} & 7.23\% & 0.0799 & 0.0838 & 4.88\% & 0.0769 & \textbf{0.0868} & 12.87\% \\
    \midrule
    \multirow{4}{*}{\textbf{Sports}} 
      & R@5  & 0.0236 & 0.0205 & 0.0233 & 0.0242 & 3.86\% & 0.0233 & 0.0240 & 3.00\% & \underline{0.0418} & \textbf{0.0612} & 46.41\% \\
      & R@10 & 0.0363 & 0.0287 & 0.0350 & 0.0384 & 9.71\% & 0.0379 & 0.0392 & 3.43\% & \underline{0.0674} & \textbf{0.0931} & 38.13\% \\
      & N@5  & 0.0157 & 0.0143 & 0.0154 & 0.0153 & -0.65\% & 0.0150 & 0.0154 & 2.67\% & \underline{0.0274} & \textbf{0.0419} & 52.92\% \\
      & N@10 & 0.0189 & 0.0170 & 0.0190 & 0.0198 & 3.13\% & 0.0197 & 0.0203 & 3.05\% & \underline{0.0361} & \textbf{0.0527} & 45.98\% \\
    \midrule
    \multirow{4}{*}{\textbf{Video \& Games}} 
      & R@5   & 0.0593 & 0.0600 & 0.0578 & 0.0604 & 4.50\% & 0.0511 & 0.0545 & 6.65\% & \underline{0.1110} & \textbf{0.1219} & 9.82\% \\
      & R@10  & 0.0920 & 0.0933 & 0.0903 & 0.0952 & 5.43\% & 0.0807 & 0.0830 & 2.85\% & \underline{0.1738} & \textbf{0.1881} & 8.23\% \\
      & N@5   & 0.0313 & 0.0321 & 0.0304 & 0.0316 & 3.95\% & 0.0334 & 0.0356 & 6.59\% & \underline{0.0731} & \textbf{0.0813} & 11.22\% \\
      & N@10 &  0.0418 & 0.0429 & 0.0409 & 0.0433 & 5.87\% & 0.0429 & 0.0448 & 4.43\% & \underline{0.0945} & \textbf{0.1041} & 10.16\% \\
    \bottomrule
    \end{tabular}%
\end{table*}

\section{Experiments}
\label{sec:experiments}

In this section, we conduct experiments to address the following research questions:

\noindent\textbf{RQ1:} How does DiffuReason perform compared with state-of-the-art latent reasoning methods across different backbones?

\noindent\textbf{RQ2:} What is the contribution of each component to the overall performance?

\noindent\textbf{RQ3:} How sensitive is DiffuReason to key hyperparameters, and what are the accuracy--efficiency trade-offs?

\noindent\textbf{RQ4:} What insights can we derive from in-depth analysis of DiffuReason?

\subsection{Experimental Setup}

\paragraph{Datasets.}
We evaluate DiffuReason on four benchmark datasets from the Amazon review collection. Three widely used datasets— Beauty, Musical Instruments, and Sports \& Outdoors—are adopted from the classic collection~\cite{he2016ups, ni2019justifying}, while Video \& Games is sourced from the Amazon Reviews 2023 library\footnote{\url{https://amazon-reviews-2023.github.io/}}. Table~\ref{tab:datasets} summarizes the statistics. For preprocessing, we consistently apply 5-core filtering across all datasets.\footnote{\url{https://amazon-reviews-2023.github.io/data_processing/5core.html}}
For Video \& Games, we treat interactions with ratings $>3$ as positive samples.
We sort each user’s interactions by timestamp and adopt a unified leave-one-out split for all datasets, using the last interaction for testing and the second-to-last for validation. To standardize training, each user’s interaction history is truncated or padded to a fixed length of 20, keeping the most recent interactions.

\paragraph{Baselines.}
To verify the effectiveness of our proposed framework, we compare DiffuReason with a diverse set of competitive baselines, ranging from sequential models to generative architectures. These methods are categorized into three groups:

\begin{itemize}
    \item \textbf{Sequential Models:} SASRec~\cite{kang2018self} is the representative ID-based model that employs unidirectional self-attention to capture sequential dependencies.

    \item \textbf{Generative \& Efficient Architectures:} TIGER~\cite{rajput2023recommender} represents items as Semantic IDs via RQ-VAE~\cite{zeghidour2021soundstream, lee2022autoregressive} and predicts item tuples auto-regressively. HSTU~\cite{zhai2024actions} is a high-performance sequential transducer using point-wise aggregation for efficient long-sequence processing. In our implementation, we augment HSTU with Semantic ID inputs (fused with ID embeddings) and a SID decoder to enable a fair comparison with generative methods (details in Appendix~\ref{app:implementation}).
    
    \item \textbf{Reasoning-Enhanced Methods:} ReaRec~\cite{tang2025think} extends SASRec with an explicit "Think-then-Act" paradigm. LARES~\cite{liu2025lares} employs depth-recurrent latent reasoning to iteratively refine user representations.
\end{itemize}

\paragraph{Evaluation Metrics.}
To ensure a fair evaluation, we adopt two widely used metrics: Recall@$K$ and NDCG@$K$ (Normalized Discounted Cumulative Gain), with $K$ set to 5 and 10. To eliminate sampling bias, we perform full ranking over the entire item set.

\subsection{Implementation Details}
All experiments are conducted on 8 NVIDIA H20 GPUs. We employ SASRec~\cite{kang2018self}, HSTU~\cite{zhai2024actions}, and TIGER~\cite{rajput2023recommender} as backbones. For ReaRec and LARES, we use their official implementations. Detailed architectures and hyperparameters are provided in Appendix~\ref{app:implementation}.

\paragraph{Training Protocols.}
SASRec and HSTU are trained for 300 epochs with batch size 512 and learning rates of $10^{-3}$ and $10^{-4}$, respectively. TIGER is trained for 200 epochs with batch size 256 and learning rate $10^{-4}$. For SID-based methods, item embeddings are generated by Qwen3-Embedding-8B~\cite{zhang2025qwen3} and quantized via RQ-VAE into 4 codebooks with 256 codewords each.

\paragraph{DiffuReason Configuration.}
The Latent Reasoning module is instantiated as a 2-layer Transformer Decoder with 8 attention heads, operating with a fixed CoT length of 3 steps. The Diffusion Refinement module employs 16 denoising steps with a cosine noise schedule. For RL Alignment, we sample 8 candidate trajectories ($G=8$) per input, utilizing Hit Rate as the reward signal. Finally, the loss weights $\alpha$ and $\beta$ are both set to 1.0.

\subsection{Main Results (RQ1)}
Table~\ref{tab:main_results} presents the overall performance comparison. The results demonstrate the effectiveness and generality of DiffuReason.

\paragraph{Performance Across Architectures.}
For the ID-based SASRec, \\ DiffuReason-S yields up to 19.64\% Recall@5 improvement on Beauty, surpassing reasoning-enhanced variants like ReaRec and LARES. We attribute this advantage to our diffusion refinement module. While ReaRec relies on deterministic reasoning where errors propagate irreversibly, DiffuReason introduces a probabilistic denoising process that provides inherent error correction, making the reasoning process more robust to noisy intermediate states. For the SID-based TIGER, DiffuReason-T achieves 5 to 8\% gains, demonstrating that this refinement mechanism also effectively denoises discrete Semantic IDs. The most substantial improvements are observed on HSTU, where DiffuReason-H delivers superior performance across most datasets, with gains reaching 46.41\% Recall@5 on Sports. The only exception is Instruments, where the gap narrows due to simpler user patterns.

 \paragraph{Model-Agnostic Improvements.}
DiffuReason consistently enhances all three backbone architectures, confirming its generality as a plug-and-play reasoning module. Across all datasets and metric configurations, DiffuReason variants outperform their vanilla counterparts in nearly all cases. This validates that the Think-then-Diffuse paradigm is orthogonal to the choice of item representation (ID-based vs.\ SID-based) and attention mechanism (quadratic vs.\ linear).

 \paragraph{Why HSTU Benefits Most.}
HSTU exhibits substantially larger gains than SASRec or TIGER (e.g., 46\% vs.\ 4\% on Sports). We hypothesize this stems from a fundamental trade-off in HSTU's design. Its point-wise aggregation mechanism sacrifices quadratic attention's expressive power for linear efficiency. While sufficient for simple patterns, this architectural constraint limits HSTU's capacity to capture complex, multi-hop dependencies in challenging scenarios. DiffuReason effectively compensates for this limitation by offloading deliberative reasoning to a dedicated module. In other words, our framework enables lightweight architectures to achieve heavyweight reasoning capabilities, combining HSTU's deployment efficiency with deep sequential understanding.

\begin{figure*}[t]
  \centering
  \includegraphics[width=0.98\linewidth]{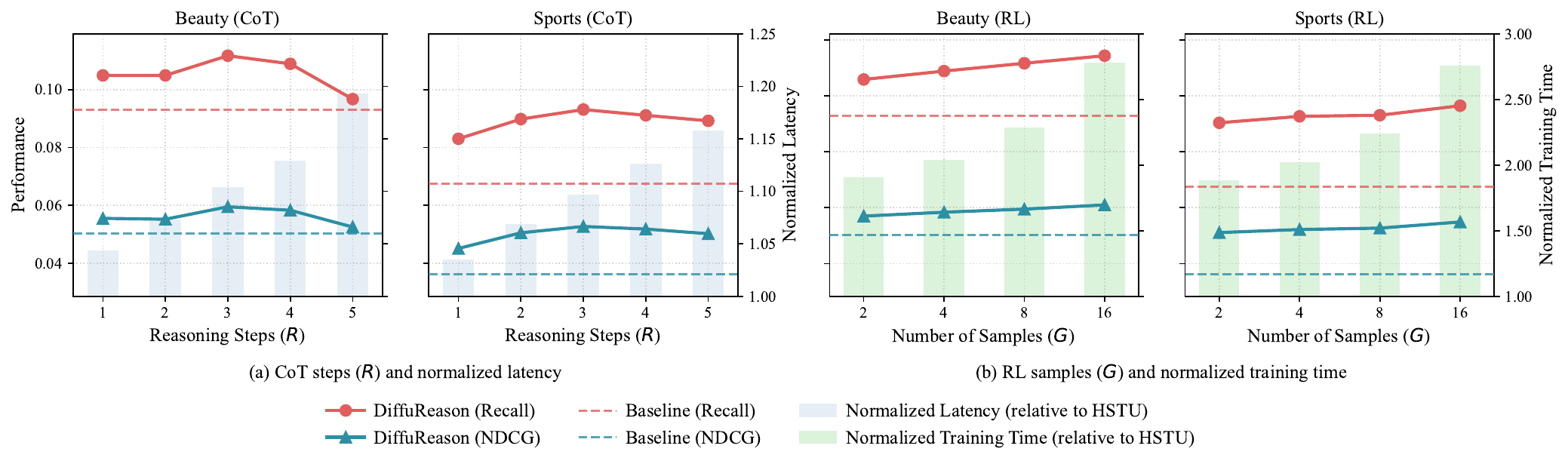}
  \vspace{-2mm}
  \caption{Efficiency analysis of DiffuReason. (a) Performance (Recall/NDCG) and normalized inference latency under different reasoning steps $R$. (b) Performance (Recall/NDCG) and normalized per-batch training time under different RL sample sizes $G$. Bars denote normalized cost relative to the HSTU backbone.}
  \label{fig:ablation_rl}
  \vspace{-3mm}
\end{figure*}

\begin{table}[t]
\centering
\caption{Ablation study of DiffuReason on two datasets. The best results are highlighted in bold. "w/o" denotes removing a specific module from the full model.}
\label{tab:ablation}
\small
\begin{tabular}{l|cc|cc}
\toprule
\multicolumn{1}{c|}{Dataset} & \multicolumn{2}{c|}{Beauty} & \multicolumn{2}{c}{Sports} \\
\multicolumn{1}{c|}{Metric} & R@10 & N@10 & R@10 & N@10 \\ 
\midrule
\textbf{DiffuReason} & \textbf{0.1117} & \textbf{0.0595} & \textbf{0.0931} & \textbf{0.0527} \\ 
\midrule
w/o CoT & 0.1013 & 0.0545 & 0.0807 & 0.0442 \\
w/o Diffusion & 0.1095 & 0.0571 & 0.0887 & 0.0496 \\
w/o RL & 0.1093 & 0.0578 & 0.0915 & 0.0510 \\
w/o MSE & 0.1115 & 0.0589 & 0.0930 & 0.0534 \\
w/o Diffusion+RL & 0.1055 & 0.0571 & 0.0831 & 0.0449 \\
\midrule
with KL & 0.1048 & 0.0558 & 0.0865 & 0.0484 \\ 
with Two-stage & 0.1099 & 0.0587 & 0.0922 & 0.0519 \\ 
\bottomrule
\end{tabular}
\vspace{-3mm}
\end{table}

\subsection{Ablation Study (RQ2)}
To rigorously validate the contribution of each component and the superiority of our training strategy, we conduct a comprehensive ablation study on the Beauty and Sports datasets. The results are reported in Table~\ref{tab:ablation}.

\paragraph{Impact of Core Components (CoT and Diffusion).}
The most significant performance degradation occurs in the w/o CoT variant, where removing the latent reasoning module leads to a sharp drop (e.g., Recall@10 drops from 0.0931 to 0.0807 on Sports). This confirms that standard implicit representations (System 1) are insufficient for capturing complex user interests, highlighting the critical necessity of explicit multi-step deduction.
Furthermore, removing the generative refinement (w/o Diffusion) also results in consistent performance losses. This validates that deterministic reasoning chains inevitably contain noise; the diffusion process is essential to probabilistically denoise the initial thoughts and refine them into precise user representations.
Notably, when both modules are removed (w/o Diffusion+RL), the performance deteriorates further, demonstrating that the synergy between reasoning and refinement is the cornerstone of DiffuReason.

\paragraph{Analysis of Optimization Objectives (MSE vs. RL)}
Comparing optimization targets reveals interesting nuances. Removing the reconstruction loss (w/o MSE) causes only minor fluctuations, whereas removing the alignment loss (w/o RL) leads to a noticeable decline. This suggests that while MSE helps stabilize the latent anchor, the GRPO-based RL alignment is the dominant driver for bridging the gap between latent generation and discrete ranking metrics.

\paragraph{End-to-End vs. Two-Stage Training.}
A critical contribution of this work is the unified end-to-end training paradigm.  We compare against a Two-stage variant that pre-trains with MSE loss for 300 epochs before RL fine-tuning for another 300 epochs. Despite the much longer training time, the Two-stage method fails to outperform our approach and yields results similar to the w/o RL baseline.

This empirical evidence directly validates our hypothesis in Section~\ref{sec:intro}: the initial decoupled stage inevitably biases the policy towards specific trajectories, thereby constraining the exploration potential of the subsequent RL optimization. Consequently, the two-stage paradigm restricts the synergistic evolution of the reasoning and refinement modules, whereas our end-to-end strategy ensures better alignment.

\paragraph{Impact of KL Divergence.}
We investigate the effect of the KL divergence constraint by adding it to the RL objective with a weight of 0.1 (with KL). Surprisingly, this leads to a significant performance drop compared to the full model. We attribute this to the fact that the base recommendation backbone is not as over-parameterized as Large Language Models. In this context, enforcing a strict KL penalty excessively limits the model's exploration capability, preventing it from discovering optimal refinement trajectories. This justifies our design choice of using the Anchor State as a natural constraint rather than an explicit KL loss.

\begin{figure}[t]
  \centering
  \includegraphics[width=0.98\linewidth]{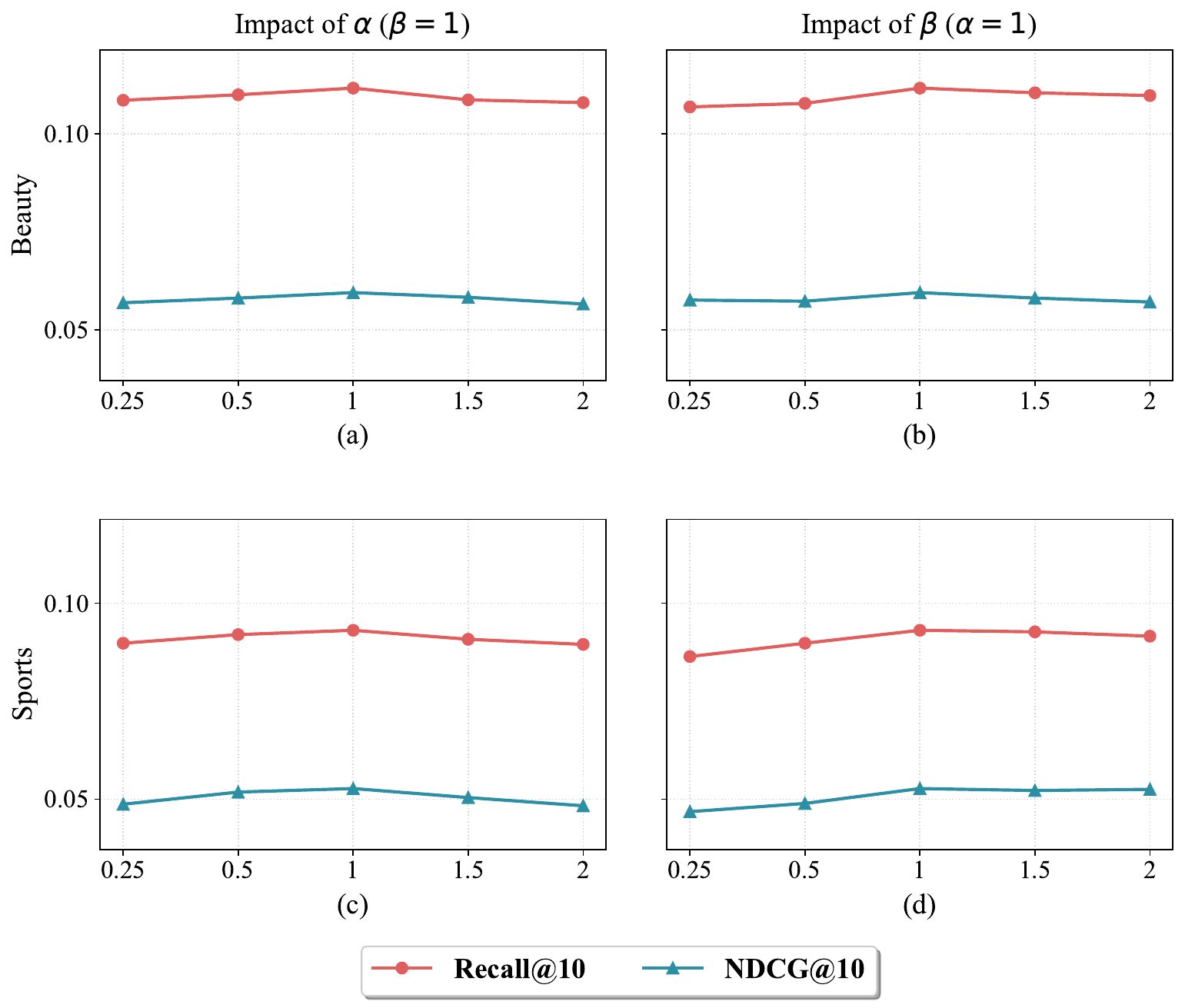}
  \caption{Sensitivity to loss weights $\alpha$ (MSE) and $\beta$ (RL) on Beauty and Sports. We vary one weight while fixing the other to 1.0, and report Recall@10 and NDCG@10.}
  \label{fig:sensitivity}
\end{figure}

\subsection{Hyperparameter Sensitivity and Efficiency (RQ3)}
We investigate the trade-off between performance and computational cost, as well as robustness to key hyperparameters.

\paragraph{Impact of Reasoning Steps $R$.}
Figure~\ref{fig:ablation_rl}(a) shows the effect of varying $R$ from 1 to 5. Performance follows an inverted U-shaped trend, peaking at $R=3$ before saturating or degrading. This suggests that while multi-step reasoning clarifies intent, excessive steps may introduce noise in the latent space. Notably, even $R=1$ already surpasses the HSTU baseline, confirming that the reasoning mechanism itself is effective regardless of depth.

At the optimal $R=3$, we observe substantial gains with modest inference overhead. On Beauty, Recall@10 improves from 0.093 to 0.1117 (+20.1\%) and NDCG@10 from 0.0502 to 0.0595 (+18.5\%). On Sports, the gains are even larger with Recall@10 improving by 38.1\% and NDCG@10 by 46.0\%. Meanwhile, inference latency increases by only around 10\%. The per-batch training time approximately doubles (2.2 to 2.3$\times$), but this is an offline cost that does not affect deployment. This favorable trade-off stems from our design of performing reasoning purely in the compact latent space, avoiding costly token-level generation.

\paragraph{Impact of RL Sample Size $G$.}
Figure~\ref{fig:ablation_rl}(b) examines $G$ ranging from 2 to 16. Performance improves monotonically within the tested range, with more pronounced gains on the complex Sports dataset. However, increasing $G$ raises training cost while inference latency remains completely unchanged.

We analyze the marginal cost-benefit to justify our default choice. Detailed efficiency statistics are provided in Appendix~\ref{app:efficiency_data}. On Beauty, increasing $G$ from 8 to 16 extends training time by 21.6\% (52.55s to 63.93s per batch) while Recall@10 improves by only 2.4\% (0.1117 to 0.1144). On Sports, the pattern is similar with 22.9\% more training time for 3.7\% performance gain. This diminishing return identifies $G=8$ as a practical sweet spot that balances offline training budget against online performance.

\paragraph{Sensitivity to Loss Weights $\alpha$ and $\beta$.}
Figure~\ref{fig:sensitivity} shows that performance remains stable across a wide range (0.25 to 2.0) for both weights, with the optimum around 1.0. We observe asymmetric sensitivity between the two parameters.

Setting $\alpha$ (MSE weight) too high causes notable degradation. For example, on Sports, Recall@10 drops from 0.0931 at $\alpha$=1 to 0.0895 at $\alpha$=2. We attribute this to over-constrained latent representations that strictly adhere to static target embeddings, hindering exploration of semantically similar but better-ranked items. In contrast, the model tolerates variations in $\beta$ (RL weight) better, with only minor fluctuations across the tested range.

These observations suggest that MSE serves as a stabilizing auxiliary objective, but excessive weight suppresses exploration. This is consistent with our RQ2 finding that RL alignment is the dominant driver for bridging latent generation and ranking metrics.

\begin{figure}[t]
  \centering
  \includegraphics[width=\linewidth]{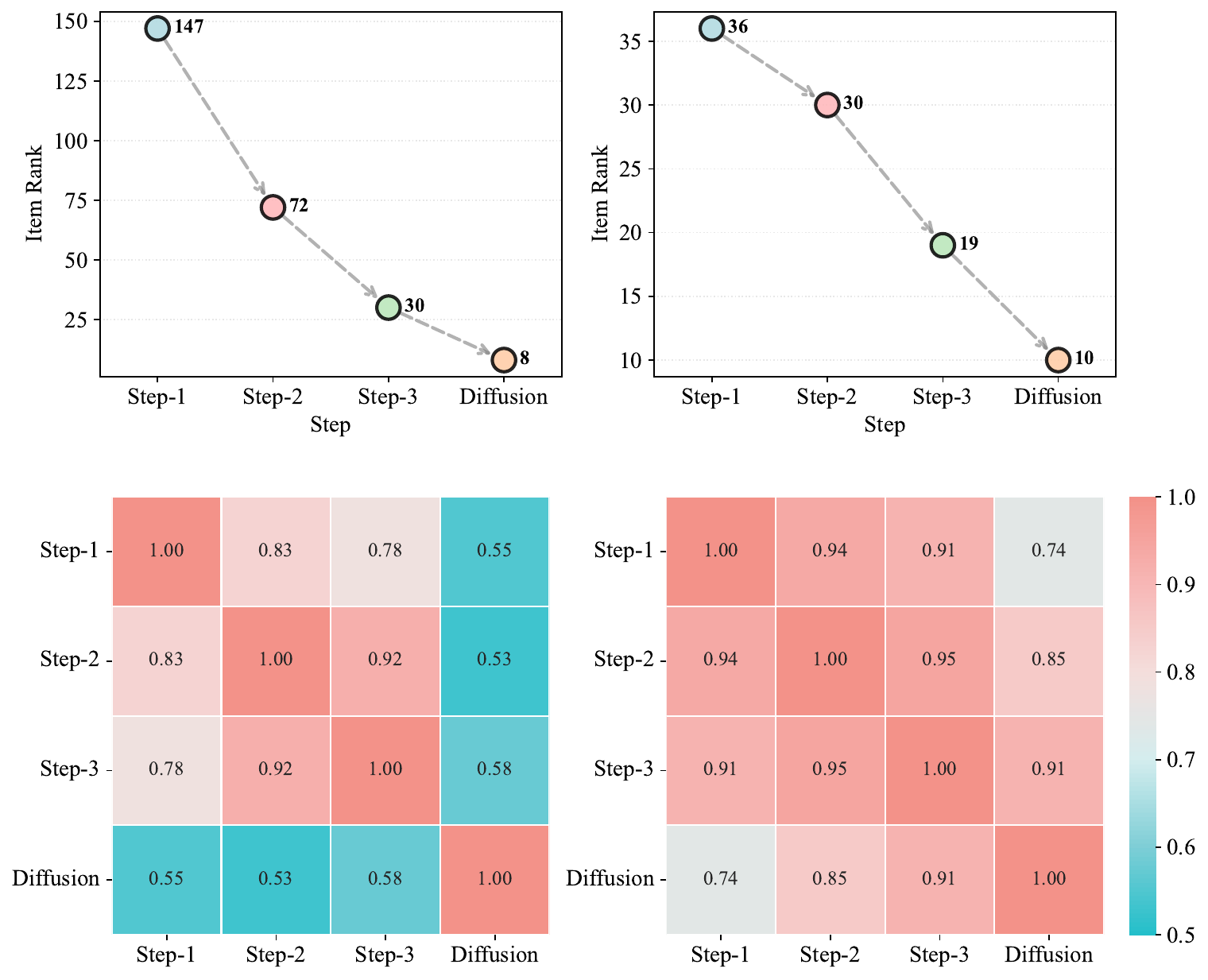}
    \caption{Visualization of the Think-then-Diffuse trajectory. Top: rank of the ground-truth item across Reasoning steps and the Diffusion stage (lower is better). Bottom: cosine similarity between latent representations at different steps.}
  \label{fig:case_study}
  \vspace{-3mm}
\end{figure}

\subsection{In-depth Analysis (RQ4)}
We provide trajectory-level visualizations and sparsity-group breakdowns to better understand DiffuReason.

\paragraph{Visualization of Reasoning Trajectory.} Figure~\ref{fig:case_study} tracks the ground-truth item rank (top) and representation similarity (bottom) along Step~1--3 and Diffusion. The rank consistently improves across Reasoning steps, while the Diffusion stage yields the largest jump (e.g., 30$\rightarrow$8), indicating that generative refinement is crucial for turning intermediate hypotheses into precise predictions. Across cases, we observe both a "repair" pattern (small CoT gains followed by a large Diffusion correction) and a "progressive" pattern (steady improvements throughout), suggesting robust behavior.

The similarity heatmaps further show that CoT representations remain highly similar across Step~1--3 (e.g., 0.78--0.95), consistent with a coherent progressive refinement. In contrast, Diffusion deviates substantially from intermediate CoT states (similarity 0.53--0.58) yet achieves the best rank. This indicates that improved decoding does not require representational proximity. The diffusion module learns to transform rather than incrementally refine the reasoning output.

\paragraph{Robustness Across Sparsity Groups.}

Following prior works~\cite{yang2023debiased, tang2025think}, we partition users into four groups by sequence length with equal-sized test sets per group (UG-0 to UG-3; higher indices indicate longer histories). For items, we split them into four popularity buckets by interaction frequency (IG-0 to IG-3); since item popularity is highly skewed, the number of test instances differs substantially across item groups (head $\gg$ tail). We therefore report relative improvements within each group for fair comparison.

\begin{figure}[t]
  \centering
  \includegraphics[width=\linewidth]{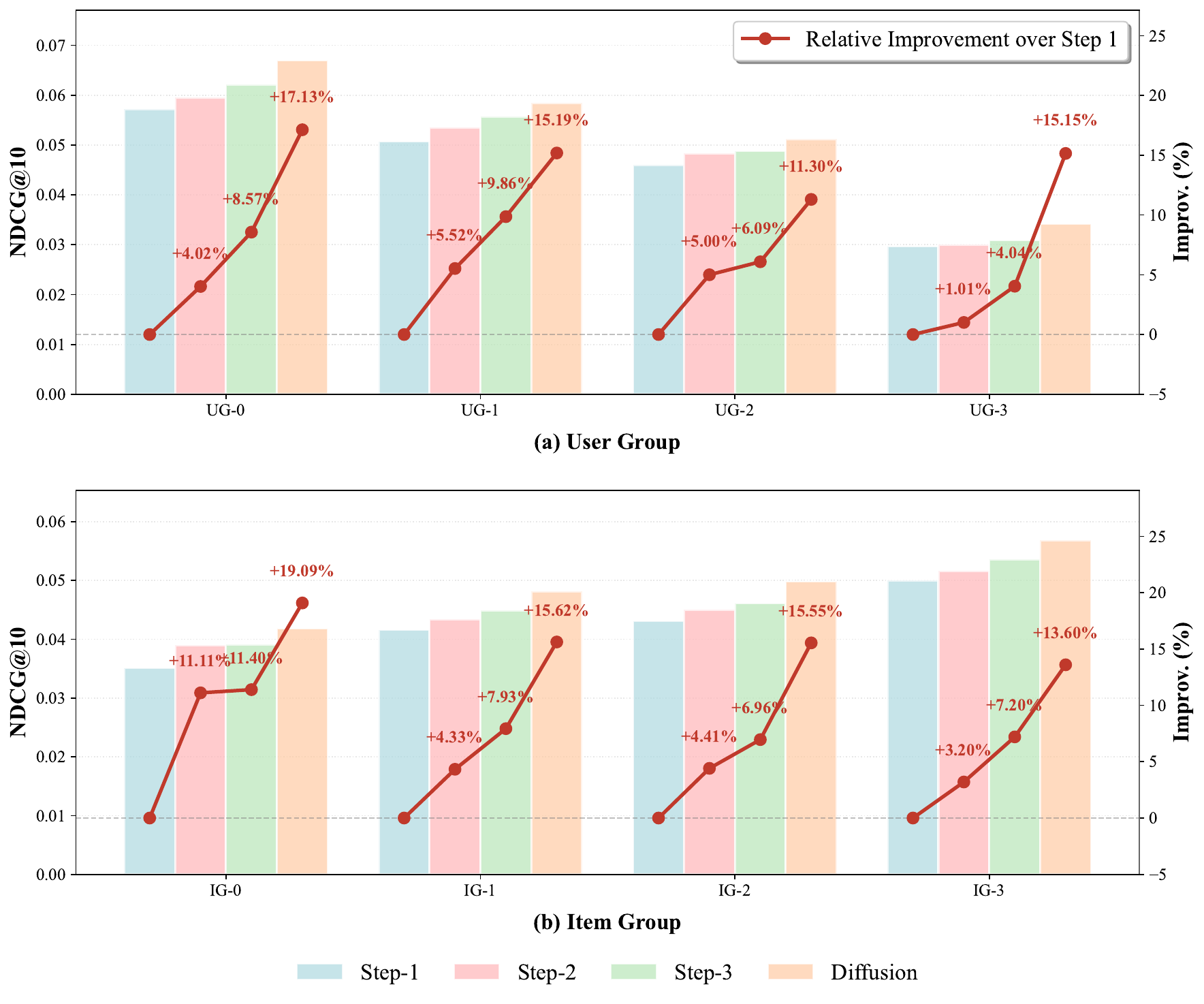}
\caption{Performance Improvement across user/item subgroups. Users are grouped by sequence length (UG-0 to UG-3), and items are grouped by interaction frequency (IG-0 to IG-3). Bars report metric values at Step~1--3 and Diffusion; the line shows relative improvement over Step~1 within each group.}
  \label{fig:sparsity_analysis}
\end{figure}

Figure~\ref{fig:sparsity_analysis} shows consistent gains from Step~1 to Diffusion across all groups. The sparse groups benefit more from deeper CoT (e.g., UG-0: +17.13\%), indicating that latent reasoning is especially helpful when behavioral evidence is limited. In dense groups, CoT brings smaller marginal gains, but Diffusion still provides a substantial additional improvement, suggesting that its main contribution shifts toward denoising and reconciling potentially conflicting interests in long histories. Consistent trends are also observed on the Beauty dataset, as detailed in Appendix~\ref{app:additional_sparsity}.

\subsection{Online A/B Testing}

To validate DiffuReason in a real-world industrial setting, we deployed it in the Tencent Ads platform for the WeChat Channels recommendation scenario. We conducted a strictly controlled online A/B test for 5 days on 20\% of production traffic.

The online baseline is a highly optimized production model with an architecture similar to HSTU. According to experiment results, DiffuReason delivers statistically significant uplifts on key business metrics: GMV (Gross Merchandise Value) increases by +0.7902\%, and Ad Consumption (revenue) increases by +1.1462\%. Given the scale and high-concurrency nature of the WeChat Channels traffic, these gains translate into substantial economic impact. Overall, the results indicate that the proposed Think-then-Diffuse paradigm not only improves offline accuracy but also generalizes well to large-scale, high-throughput online recommendation systems.

\section{Conclusion}
\label{sec:conclusion}

In this work, we presented DiffuReason, a unified framework that enhances sequential recommendation by combining latent reasoning with generative refinement. The framework generates multi-step Thinking Tokens to capture complex user intent, applies diffusion-based denoising to refine intermediate representations, and employs GRPO alignment to bridge continuous generation with discrete ranking objectives. A key contribution is our end-to-end training paradigm that allows all components to co-evolve, mitigating the misalignment commonly introduced by decoupled two-stage pipelines. Experiments on four benchmarks demonstrate consistent improvements across diverse backbones, with gains up to 46\%. Online A/B tests on WeChat Channels further validate its industrial applicability.  Promising directions for future work include adaptive reasoning that tailors CoT depth to query complexity, and expansion into multi-modal scenarios incorporating visual and textual contexts. Exploring more expressive generative architectures is also worth investigating.

%% file: paragraphs/5.Appendix.tex
\newpage
\section{Experimental Setup and Hyperparameter}~\label{append:hyper}

\begin{table}[t]
\centering
\small
\caption{RQVAE hyperparameters for Semantic ID quantization.}
\label{tab:rqvae_hparams}
\begin{tabular}{ll}
\toprule
\textbf{Component} & \textbf{Setting} \\
\midrule
Latent dim ($e\_dim$) & 32 \\
\# quantizers (levels) & 4 \\
Codebook sizes ($num\_emb\_list$) & [256, 256, 256, 256] \\
Reconstruction loss & MSE (weight=1.0) \\
Optimizer & Adam (lr=$1\text{e-}3$, weight\_decay=$1\text{e-}4$) \\
Batch size / epochs & 1024 / 20000 \\
\bottomrule
\end{tabular}
\end{table}

\begin{figure}[t]
    \centering
    \includegraphics[width=\linewidth]{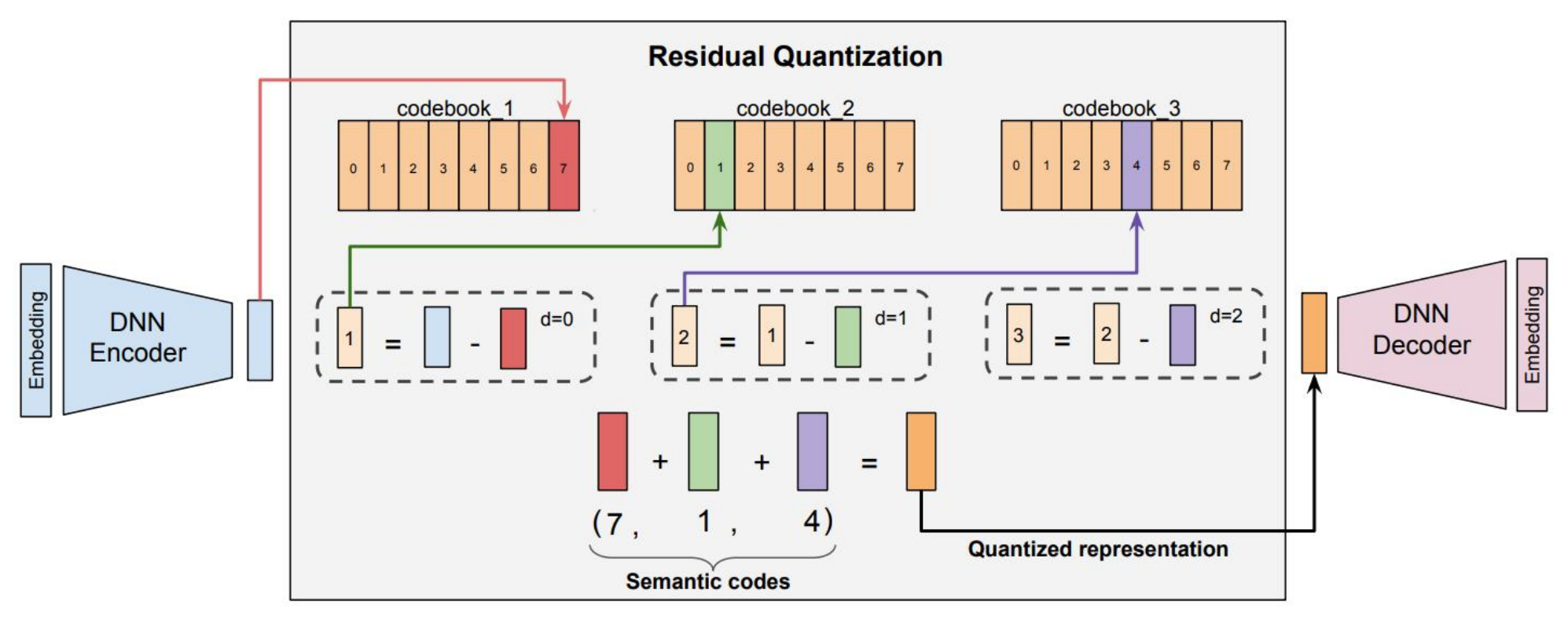}
    \caption{Overview of RQ-VAE with residual quantization (original figure from ~\cite{rajput2023recommender}).}
    \label{fig:rqvae}
\end{figure}

\paragraph{RQ-VAE for Semantic IDs (RQVAE)}
We adopt an RQ-VAE tokenizer to discretize continuous item embeddings into compact Semantic IDs (SIDs), as illustrated in Fig.~\ref{fig:rqvae} (original figure from~\cite{rajput2023recommender}).
Instead of assigning a single code, RQ-VAE generates an $M$-level SID by sequentially quantizing a latent vector with a stack of codebooks in a residual (coarse-to-fine) manner.
The SID is the concatenation of selected code indices across levels (one per codebook). When multiple items share the same SID prefix, we optionally append an additional disambiguation token to preserve a one-to-one mapping.
We summarize the RQVAE configuration used for Semantic ID quantization in Table~\ref{tab:rqvae_hparams}.

\noindent\textbf{Residual quantization.}
Let $\mathbf{v}\in\mathbb{R}^{d_{\text{in}}}$ be the (continuous) item embedding and $\mathbf{y}=\mathrm{Enc}_{\phi}(\mathbf{v})\in\mathbb{R}^{d_{e}}$ be its latent representation.
RQ-VAE employs $M$ codebooks $\{\mathcal{E}^{(m)}\}_{m=1}^{M}$, where $\mathcal{E}^{(m)}=\{\mathbf{e}^{(m)}_{k}\}_{k=1}^{K_m}$ and $\mathbf{e}^{(m)}_{k}\in\mathbb{R}^{d_e}$.
Initialize the residual as $\boldsymbol{\rho}^{(1)}=\mathbf{y}$. For $m=1,\ldots,M$, we select the nearest codeword:
\begin{equation}
\label{eq:rqvae_residual}
\begin{aligned}
j_m
&=\arg\min_{k\in\{1,\ldots,K_m\}}
\left\|\boldsymbol{\rho}^{(m)}-\mathbf{e}^{(m)}_{k}\right\|_2^2, \\
\mathbf{q}^{(m)}
&=\mathbf{e}^{(m)}_{j_m}, \\
\boldsymbol{\rho}^{(m+1)}
&=\boldsymbol{\rho}^{(m)}-\mathbf{q}^{(m)}.
\end{aligned}
\end{equation}

The quantized latent is $\hat{\mathbf{y}}=\sum_{m=1}^{M}\mathbf{q}^{(m)}$, and the decoder reconstructs $\hat{\mathbf{v}}=\mathrm{Dec}_{\psi}(\hat{\mathbf{y}})$.
The resulting SID is $\mathbf{s}=(j_1,\ldots,j_M)$ (optionally with an appended disambiguation token).

\noindent\textbf{Tokenization objective.}
We jointly train the encoder, decoder, and codebooks with the following tokenization loss:
\begin{equation}
\label{eq:rqvae_loss}
\begin{aligned}
\mathcal{L}_{\text{tok}}(\mathbf{v})
&=
\underbrace{\left\|\hat{\mathbf{v}}-\mathbf{v}\right\|_2^2}_{\text{reconstruction}}
+
\sum_{m=1}^{M}
\Big(
\underbrace{\left\|\mathrm{sg}\!\left[\boldsymbol{\rho}^{(m)}\right]-\mathbf{q}^{(m)}\right\|_2^2}_{\text{codebook}}
\\
&\quad\quad\quad\quad
+
\lambda_{\text{cm}}\,
\underbrace{\left\|\boldsymbol{\rho}^{(m)}-\mathrm{sg}\!\left[\mathbf{q}^{(m)}\right]\right\|_2^2}_{\text{commitment}}
\Big).
\end{aligned}
\end{equation}

where $\mathrm{sg}[\cdot]$ denotes the stop-gradient operator and $\lambda_{\text{cm}}$ controls the commitment strength (Table~\ref{tab:rqvae_hparams}).

\subsection{Backbone adaptation and DiffuReason integration}
\label{app:implementation}
\paragraph{SASRec baseline.}
SASRec is a discriminative Transformer encoder that directly predicts the next item via a softmax classifier over the full item set.
We use the standard SASRec configuration (embedding dimension 256, 2 Transformer layers with 2 attention heads, FFN hidden size 300, dropout 0.5, GELU activation),
trained with CrossEntropyLoss (full softmax).

\paragraph{SASRec + DiffuReason.}
For SASRec, DiffuReason is implemented as an outer-loop refinement that does not alter the SASRec backbone. Concretely, given the same input sequence, we perform multiple forward passes (rollouts) of SASRec. A Latent Reasoning module (Transformer decoder) produces a fixed-step latent CoT trajectory that conditions on the SASRec sequence representation. A diffusion model refines the latent trajectory for $T{=}16$ denoising steps (cosine schedule), and the refined latent is used to update the next-step prediction in subsequent rollouts. For RL alignment, we sample $G{=}8$ candidate trajectories per input and optimize the Hit-Rate-based reward (measured on the SASRec prediction outcomes).

\paragraph{HSTU baseline with Semantic IDs.}
The original HSTU predicts item IDs directly. To enable generative retrieval and ensure fair comparison to generative backbones, we augment HSTU with hierarchical Semantic IDs (SIDs) and an autoregressive SID decoder.
Each item is associated with a 4-level SID code (256 codes per level). For each interaction token, we embed the item ID and the 4 SID codes using separate embedding tables and sum them to obtain the final input token embedding to the HSTU encoder. The SID decoder then autoregressively predicts the 4-level SID sequence.

\paragraph{HSTU + DiffuReason.}
DiffuReason is attached on top of the SID-enabled HSTU.
The Latent Reasoning module (Transformer decoder) attends to the HSTU encoder outputs to generate a fixed-step latent CoT trajectory; the diffusion model refines the latent trajectory, and the refined latent is provided as additional conditioning context for SID prediction.
RL alignment follows the same protocol: we sample $G{=}8$ candidate trajectories and use Hit Rate as the reward signal.

\paragraph{TIGER baseline.}
TIGER follows a T5-style encoder--decoder architecture and autoregressively generates discrete SIDs.
We use the standard configuration: 4-layer encoder and 4-layer decoder, $d_{\text{model}}{=}128$, $d_{\text{ff}}{=}1024$, 6 attention heads, and dropout 0.1.

\paragraph{TIGER + DiffuReason.}
DiffuReason is integrated by conditioning the Latent Reasoning module on TIGER encoder states to produce latent CoT tokens, refining them through diffusion,
and injecting the refined latent tokens as extra conditioning signals for TIGER decoding (e.g., as additional cross-attention context).
RL alignment samples $G{=}8$ candidate trajectories per input and optimizes Hit Rate reward.

\subsection{Hyperparameters}
\label{app:hparams}

\begin{table}[t]
\centering
\small
\caption{Hyperparameters for backbones and DiffuReason. DiffuReason settings are shared across backbones unless stated otherwise.}
\label{tab:overall_hparams}
\resizebox{\linewidth}{!}{%
\begin{tabular}{lccc}
\toprule
\textbf{Setting} & \textbf{SASRec} & \textbf{HSTU} & \textbf{TIGER} \\
\midrule
\multicolumn{4}{l}{\textit{Backbone architecture}} \\
Embedding dim & 256 & 50 & 128 \\
Encoder layers & 2 & 2 & 4 \\
Attention heads & 2 & 1 & 6 \\
FFN hidden dim & 300 & -- & 1024 \\
Dropout & 0.5 & 0.2 & 0.1 \\
Activation & GELU & -- & ReLU (FFN) \\
\midrule
\multicolumn{4}{l}{\textit{Backbone training}} \\
Epochs & 300 & 200 & 200 \\
Batch size & 512 & 512 & 256 \\
Learning rate & $10^{-3}$ & $10^{-4}$ & $10^{-4}$ \\
Optimizer & Adam & Adam & Adam \\
\midrule
\multicolumn{4}{l}{\textit{SID / tokenizer}} \\
\# codebooks ($M$) & -- & 4 & 4 \\
Codebook size & -- & 256/level & 256/level \\
Tokenizer & -- & RQ-VAE (Tab.~\ref{tab:rqvae_hparams}) & RQ-VAE (Tab.~\ref{tab:rqvae_hparams}) \\
\midrule
\multicolumn{4}{l}{\textit{DiffuReason (shared)}} \\
Latent Reasoning decoder layers & \multicolumn{3}{c}{2} \\
Latent Reasoning attention heads & \multicolumn{3}{c}{8} \\
CoT length (steps) & \multicolumn{3}{c}{3} \\
Diffusion denoising steps ($T$) & \multicolumn{3}{c}{16} \\
Noise schedule & \multicolumn{3}{c}{cosine} \\
RL samples per input ($G$) & \multicolumn{3}{c}{8} \\
Reward & \multicolumn{3}{c}{Hit Rate} \\
Loss weights & \multicolumn{3}{c}{$\alpha=1.0,\ \beta=1.0$} \\
\bottomrule
\end{tabular}
}
\end{table}

Table~\ref{tab:overall_hparams} summarizes the backbone-specific configurations and the shared DiffuReason settings used throughout our experiments.
Unless stated otherwise, we keep the DiffuReason modules identical across backbones (latent reasoning depth/heads, CoT length, diffusion steps and schedule, RL sampling count, and loss weights) to ensure a controlled comparison.

\paragraph{Backbone training.}
SASRec is trained with a standard full-softmax cross-entropy objective over item IDs.
For HSTU and TIGER, we adopt SID-based prediction for generative retrieval: item embeddings are quantized into a 4-level Semantic ID (256 codes per level) using an RQ-VAE tokenizer (Table~\ref{tab:rqvae_hparams} and Fig.~\ref{fig:rqvae}).
HSTU is augmented with SID embeddings (summed with the item-ID embedding) and an autoregressive SID decoder, while TIGER follows its native T5-style encoder--decoder SID generation.

\paragraph{DiffuReason settings.}
The latent reasoning module is instantiated as a lightweight Transformer decoder that produces a fixed-length latent trajectory (3 steps) conditioned on backbone sequence representations.
The diffusion refinement performs $T{=}16$ denoising steps with a cosine noise schedule.
For RL alignment, we sample $G{=}8$ candidate trajectories per input and use Hit Rate as the reward signal; the weighting coefficients are set to $\alpha{=}1.0$ and $\beta{=}1.0$.

\section{Additional Robustness Analysis}
\label{app:additional_sparsity}

\begin{figure}[h]
  \centering
  \includegraphics[width=\linewidth]{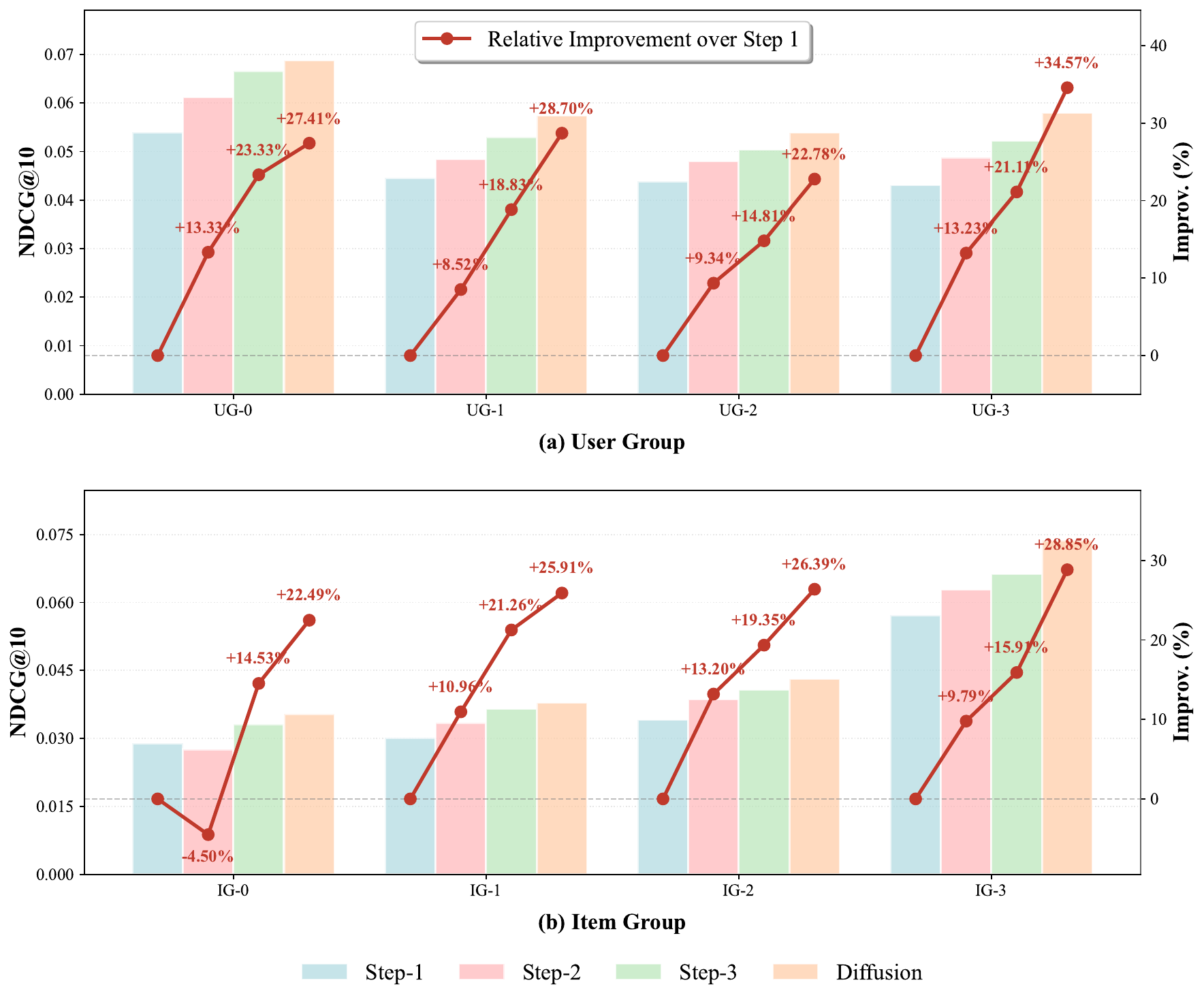}
  \caption{Performance Improvement across user and item subgroups on the Beauty dataset. The framework shows consistent gains across all user groups. Notably, in the sparse item group (IG-0), the Diffusion module effectively corrects the reasoning error observed in Step-2, validating its robustness.}
  \label{fig:beauty_sparsity}
\end{figure}

To comprehensively validate the robustness of DiffuReason, we present the performance breakdown across user and item subgroups for the Beauty dataset. The detailed results are illustrated in Figure~\ref{fig:beauty_sparsity}.

The overall trends align well with the observations on the Sports dataset discussed in the main text. Regarding user groups based on sequence length, the framework demonstrates universal effectiveness. As shown in Figure~\ref{fig:beauty_sparsity}(a), both the sparsest group UG-0 and the densest group UG-3 achieve substantial relative improvements of 27.41 percent and 34.57 percent respectively. This indicates that the Think-then-Diffuse paradigm effectively captures user intent regardless of interaction history length.

A unique and significant phenomenon emerges in the item popularity analysis presented in Figure~\ref{fig:beauty_sparsity}(b). In the long-tail group IG-0, we observe a performance dip during the intermediate CoT steps where Step-2 drops by 4.50 percent compared to Step-1. This fluctuation likely stems from the scarcity of collaborative signals for tail items, which can mislead the autoregressive reasoning process and cause error propagation. However, the subsequent Diffusion stage successfully rectifies this deviated trajectory, ultimately boosting the final performance to a 22.49 percent relative improvement. This specific case strongly corroborates the critical role of the diffusion module as a robust denoiser. It demonstrates that our generative refinement mechanism can effectively compensate for potential instabilities in the pure reasoning chain, ensuring reliable recommendations even for challenging long-tail items.

\section{Detailed Efficiency and Sensitivity Data}
\label{app:efficiency_data}

To supplement the normalized visualization in Figure~\ref{fig:ablation_rl} of the main text, we provide the detailed numerical results in Table~\ref{tab:detailed_efficiency}. This table reports the absolute training time (ms per batch), total inference time on the full test set (seconds), and the corresponding recommendation performance (Recall@10, NDCG@10) for both Beauty and Sports datasets. The baseline model corresponds to the setting with CoT=0 and RL=0 (Vanilla HSTU).

Analyzing the exact values confirms our efficiency claims. Regarding the reasoning steps $R$, increasing $R$ from 0 (baseline) to 3 incurs a manageable increase in inference latency. For instance, on the Beauty dataset, the total inference time rises from 21.81s to 24.08s, representing a modest 10.4 percent overhead. Similarly, on Sports, the time increases from 34.39s to 37.71s, a 9.6 percent increase. This verifies that the "Think-then-Diffuse" mechanism improves performance significantly (e.g., Recall@10 from 0.0930 to 0.1117 on Beauty) with only a marginal cost in online serving latency.

Regarding the RL sample size $G$, the data highlights the decoupling of training and inference costs. While increasing $G$ from 2 to 16 significantly raises the training budget (e.g., from 67.86ms to 99.36ms per batch on Sports), the inference time remains strictly constant at 37.71s. This validates that the complexity of the RL alignment phase is entirely absorbed offline, imposing zero additional burden on the online inference system.

\begin{table}[h]
\centering
\caption{Detailed performance and efficiency statistics. ``Time (Train)'' denotes the training time per batch (ms), and ``Time (Inf.)'' denotes the total inference time for the full test set (s). The Baseline corresponds to HSTU (CoT=0, RL=0). All other variants use fixed parameters when not being varied (default CoT=3, RL=8).}
\label{tab:detailed_efficiency}
\resizebox{\linewidth}{!}{
\begin{tabular}{cc|cccc|cccc}
\toprule
\multicolumn{2}{c|}{Setting} & \multicolumn{4}{c|}{Beauty Dataset} & \multicolumn{4}{c}{Sports Dataset} \\
\cmidrule(lr){1-2} \cmidrule(lr){3-6} \cmidrule(lr){7-10}
CoT Steps & RL Samples & Time (Train) & Time (Inf.) & R@10 & N@10 & Time (Train) & Time (Inf.) & R@10 & N@10 \\
\midrule
\multicolumn{10}{c}{\textit{Baseline (HSTU)}} \\
0 & 0 & 22.98 & 21.81 & 0.0930 & 0.0502 & 36.03 & 34.39 & 0.0674 & 0.0361 \\
\midrule
\multicolumn{10}{c}{\textit{Impact of Reasoning Steps ($R$) with fixed RL ($G=8$)}} \\
1 & 8 & 39.79 & 22.77 & 0.1049 & 0.0555 & 62.04 & 35.59 & 0.0830 & 0.0451 \\
2 & 8 & 46.38 & 23.40 & 0.1049 & 0.0552 & 72.12 & 36.76 & 0.0898 & 0.0505 \\
\textbf{3} & \textbf{8} & 52.55 & 24.08 & \textbf{0.1117} & \textbf{0.0595} & 80.82 & 37.71 & \textbf{0.0931} & \textbf{0.0527} \\
4 & 8 & 59.23 & 24.62 & 0.1089 & 0.0583 & 91.67 & 38.73 & 0.0911 & 0.0518 \\
5 & 8 & 64.42 & 26.03 & 0.0967 & 0.0525 & 101.56 & 39.82 & 0.0892 & 0.0502 \\
\midrule
\multicolumn{10}{c}{\textit{Impact of RL Samples ($G$) with fixed CoT ($R=3$)}} \\
3 & 2 & 43.79 & 24.08 & 0.1059 & 0.0570 & 67.86 & 37.71 & 0.0904 & 0.0511 \\
3 & 4 & 46.80 & 24.08 & 0.1089 & 0.0584 & 72.87 & 37.71 & 0.0927 & 0.0522 \\
3 & 8 & 52.55 & 24.08 & 0.1117 & 0.0595 & 80.82 & 37.71 & 0.0931 & 0.0527 \\
3 & 16 & 63.93 & 24.08 & 0.1144 & 0.0610 & 99.36 & 37.71 & 0.0965 & 0.0549 \\
\bottomrule
\end{tabular}
}
\end{table}

\subsection{Diffusion Model Details}
\label{app:diffusion}

\subsubsection{Background: Denoising Diffusion Probabilistic Models}

Our generative refinement module is inspired by DDPM~\cite{ho2020denoising} in the sense of performing iterative denoising in a Markovian manner. However, we do not instantiate the standard DDPM forward process $q(x_t|x_0)$ nor use the analytic posterior mean derived from $(\beta_t, \bar{\alpha}_t)$. Instead, consistent with the main text, we warm-start the trajectory from the final thinking token and directly parameterize a conditional reverse transition with a learned diagonal Gaussian at each step.

\textbf{Reverse Process.} At each denoising step $t$, the denoiser outputs a diagonal Gaussian:
\begin{equation}
p_\phi(x^{t-1}\mid x^{t}, c)=\mathcal{N}\!\left(x^{t-1};\,\boldsymbol{\mu}_\phi(x^{t},t,c),\,\mathrm{diag}\!\left(\boldsymbol{\sigma}_\phi(x^{t},t,c)^2\right)\right).
\label{eq:app_reverse}
\end{equation}

\subsubsection{Implementation Details of the Refinement Module}

We parameterize each reverse transition with a learned diagonal Gaussian in the latent space. Noise is applied independently to each token position.

\textbf{Noise Schedule.} We use a cosine schedule~\cite{nichol2021improved} with $T{=}16$ steps:
\begin{equation}
\bar{\alpha}_t = \cos^2\!\left(\frac{t/T + s}{1 + s} \cdot \frac{\pi}{2}\right), \quad s = 0.008
\end{equation}

\textbf{Denoising Network.} The denoiser is a 3-layer MLP with SiLU activations. It takes the concatenation of the condition vector $c$ (Eq.~2) and the noised state with time embedding as input, i.e., $[c,\, x_t + t_{\text{emb}}] \in \mathbb{R}^{2d}$. The hidden layers map $2d \!\to\! 4d \!\to\! 4d$ with dropout (rate 0.1), and the output layer produces $2d$ dimensions that are split into mean $\boldsymbol{\mu}$ and log-std $\log\boldsymbol{\sigma}$. The standard deviation is constrained via $\boldsymbol{\sigma} = \mathrm{softplus}(\log\boldsymbol{\sigma}) + 10^{-5}$ and clamped to $[10^{-3}, 1.0]$.

\textbf{Time Embedding.} Diffusion timesteps are encoded using sinusoidal positional encoding followed by a two-layer MLP with SiLU activation, mapping scalar $t$ to a $d$-dimensional vector $t_{\text{emb}}$.

\textbf{Inference.} At test time, we run the full $T$-step reverse process starting from $x^T = \tau_R + \epsilon_{\text{init}}$ (Eq.~3) and use the predicted mean $\boldsymbol{\mu}$ as the deterministic anchor for decoding. No stochastic sampling is performed during inference.

%
%
%
%


\subsection{Adapting GRPO to Continuous Latent Space}
\label{app:grpo}

Standard GRPO~\cite{shao2024deepseekmath} is designed for discrete token generation in large language models, where the policy probability $\pi_\theta(y|x) = \prod_j \pi_\theta(y_j | x, y_{<j})$ factorizes over discrete tokens. In our setting, the ``action'' is a continuous latent vector $z_i \in \mathbb{R}^d$ sampled from the diffusion output distribution, which requires a different formulation of the importance sampling ratio $\rho_i$. We detail this adaptation below.

\subsubsection{Gaussian Policy}

As described in Section~\ref{sec:diffusion}, the diffusion refinement module outputs the parameters of a diagonal Gaussian distribution:
\begin{equation}
\pi_\theta(z) = \mathcal{N}\!\left(z;\, \boldsymbol{\mu},\, \mathrm{diag}(\boldsymbol{\sigma}^2)\right)
\end{equation}
where $\boldsymbol{\mu}, \boldsymbol{\sigma} \in \mathbb{R}^d$ are the mean and standard deviation predicted by the denoising network conditioned on the input sequence. Each candidate $z_i = \boldsymbol{\mu} + \boldsymbol{\sigma} \odot \epsilon_i$ with $\epsilon_i \sim \mathcal{N}(0, \mathbf{I})$ is a sample from this policy.

\subsubsection{Importance Sampling Ratio}

The importance sampling ratio $\rho_i$ in Eq.~\ref{eq:loss_align} measures the density change of candidate $z_i$ between the current policy $\pi_\theta$ and the old policy $\pi_{\theta_{\text{old}}}$ used to generate the samples:
\begin{equation}
\rho_i = \frac{\pi_\theta(z_i)}{\pi_{\theta_{\text{old}}}(z_i)}
\label{eq:rho}
\end{equation}
Since the policy is a diagonal Gaussian, the log-density has a closed-form expression:
\begin{equation}
\log \pi_\theta(z_i) = -\frac{1}{2}\sum_{j=1}^{d}\left[\frac{(z_{i,j} - \mu_j)^2}{\sigma_j^2} + \log \sigma_j^2\right] - \frac{d}{2}\log(2\pi)
\label{eq:logprob}
\end{equation}
The ratio is then computed in log-space for numerical stability:
\begin{equation}
\rho_i = \exp\!\Big(\log\pi_\theta(z_i) - \log\pi_{\theta_{\text{old}}}(z_i)\Big)
\label{eq:rho_exp}
\end{equation}
Note that the constant term $-\frac{d}{2}\log(2\pi)$ cancels in the subtraction. In practice, we detach $\pi_{\theta_{\text{old}}}$ (i.e., compute its log-density with stopped gradients) so that gradients flow only through $\pi_\theta$.

